\documentclass[]{article}
\usepackage{graphicx}                    
\usepackage[square]{natbib}
\begin{document}
\title{Type II Shocks Characteristics: Comparison with associated CMEs and Flares}

\author{G. Pothitakis, E. Mitsakou, P. Preka--Papadema, X. Moussas, C. Caroubalos\\University of Athens, 15783 Athens, Greece\\
C. E. Alissandrakis\\University of Ioannina, 45110 Ioannina, Greece\\
A. Hillaris\\University of Athens, 15783 Athens, Greece\\
P. Tsitsipis, A. Kontogeorgos\\Technological Education Institute of Lamia, Lamia, Greece\\
J.-L. Bougeret, G. Dumas\\}
\maketitle
\begin{abstract}
A number of metric (100-650 MHz) typeII bursts was recorded by the ARTEMIS-IV radiospectrograph 
in the 1998-2000 period; the sample includes both CME driven shocks and shocks originating from flare blasts. 
We study their characteristics in comparison with characteristics of associated CMEs and flares.
\end{abstract}

\section{The relationship of type II bursts with CMEs and flares.}

\begin{figure}
  \centerline{\includegraphics[width=\textwidth,height=6cm]{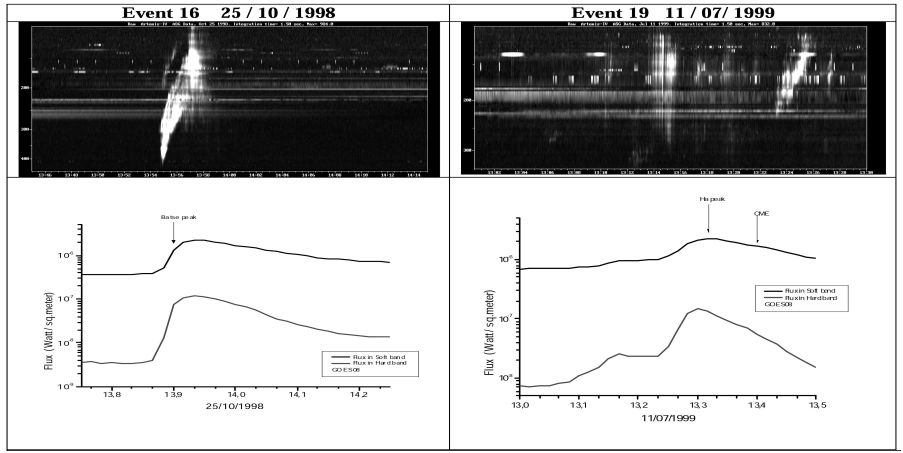}}
  \caption{ARTEMIS--IV Dynamic Spectra (Top) and light curves of the associated GOES 08 SXR Flares (Bottom).
Left Panels: The Non--CME type II burst of the 25th October 1998. right Panes: The CME--associated 
type II burst of the 11th June 1999. The latter is pereceded by a type III group.}
\label{DSpectra}
\end{figure}
Type II bursts are radio signatures of MHD shock waves in the solar corona; they either orginate 
from flare blasts or are driven by CMEs. Although the CME driven type II scenario has been widely accepted for the 
Interplanetary Shocks the issue is still open for their coronal counterparts (cf. for example \cite{Kahler}). 
The CMEs, on the other hand, are energetic phenomena consisting of mass and \emph{frozen in} magnetic field 
expulsions from the Sun. They, more often than not, coincide with flares, although the cause and effect
relationship is, as yet, obscure (cf. \cite{Harrison}). 
A thorough study of coronal bursts (in the metric wavelengths) with associated 
activity such as flares and CMEs may contribute to the resolution of these questions.
\begin{figure}
  \centerline{\includegraphics[width=\textwidth,height=6cm]{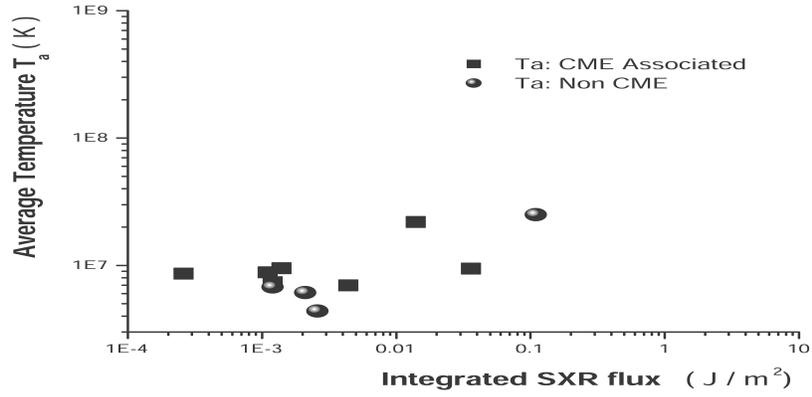}}
  \caption{Average temperature (Ta) versus SXR integrated flux for the eleven events of our data set. 
CME associated events are squares while non--CME events are dots.}
\label{FlareT}
\end{figure}
To this end, in this report, we examine a relatively small sample of type II bursts and compare their 
characteristics with these of associated CMEs and flares; it is, in fact, a three fold comparison
of type II--SXR flare, type II--CME and flare--CME.

Our data set consists of eleven type II bursts accompanied by SXR flares; 
seven of  them were associated with a CME while the remaining four were characterised as Non--CME events.
\begin{figure}
\begin{minipage}[t]{6cm}
\begin{center}
\includegraphics[width=6cm,height=6cm]{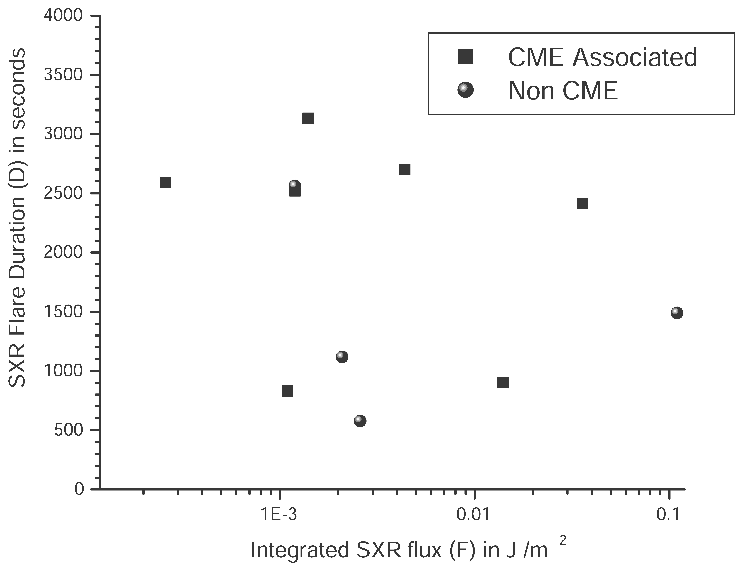}
\end{center}
\end{minipage}
\hfill
\begin{minipage}[t]{6cm}
\begin{center}
\includegraphics[width=6cm,height=6cm]{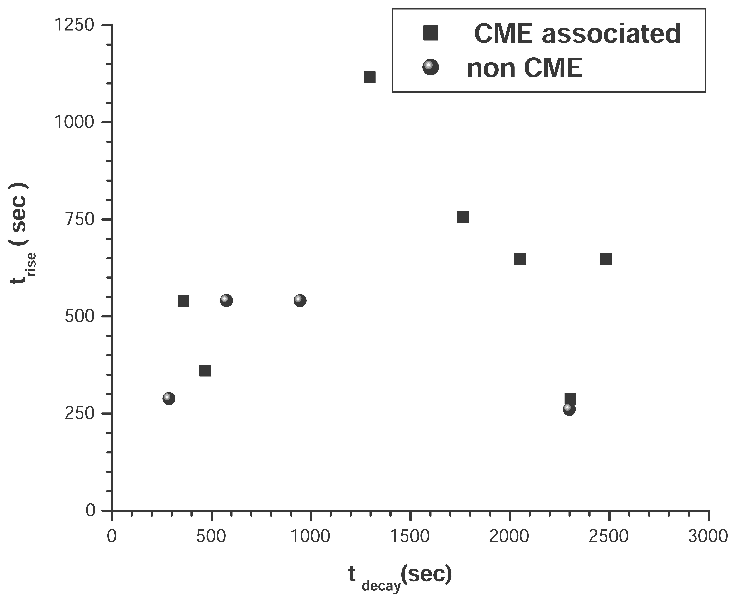}
\end{center}
\end{minipage}
  \caption{Comparison of parameter relationship  between CME and Non--CME events; 
CME associated events are squares while non--CME events are dots.Left panel:Flare duration (D) 
versus SXR integrated flux (F). Right panel:Flare rise time versus decay time.}
\label{FlareProperties}
\end{figure}
The type II radio bursts were recorded in the 100--650 MHz range by the ARTEMIS--IV 
radiospectrograph\footnote{www.uoa.gr/~artemis} (\cite{Caroubalos}) in the 1998-2000 period.
A cataloque has been published in \cite{Caroubalos04} where the characteristics of these bursts, such
as type II radial velocity, frequency range, duration, associated activity, etc. are reported. 
The associated CMEs were obtained from SOHO/LASCO lists\footnote{cdaw.gsfc.nasa.gov/CME\_list} 
(cf. \cite{Yashiro}); 
GOES SXR and Ha flare data were found in the NGDC\footnote{www.ngdc.noaa.gov/stp/SOLAR/}. Two examples 
of type II dynamic spectra and the associated GOES SXR light curves are demonstrated in 
figure \ref{DSpectra}.
\begin{figure}
\begin{minipage}[t]{6cm}
\begin{center}
\includegraphics[width=6cm,height=6cm]{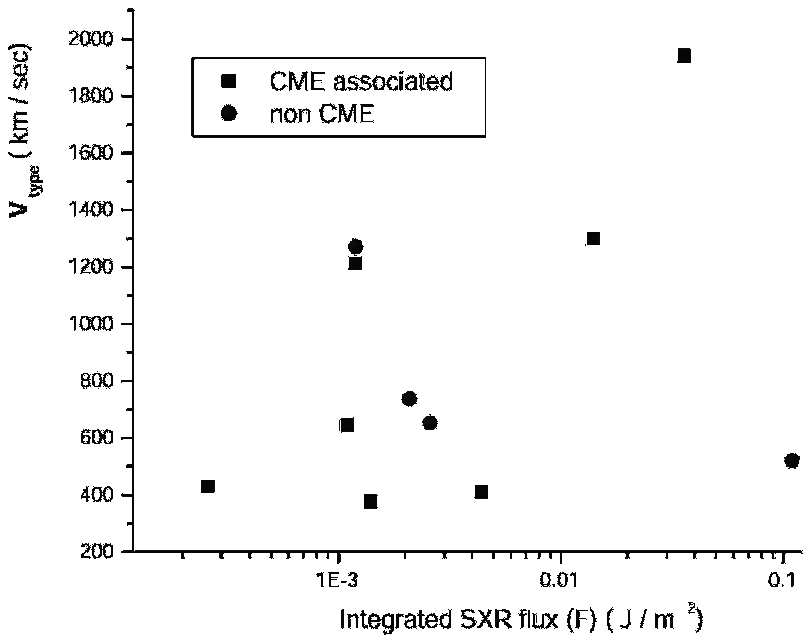}
\end{center}
\end{minipage}
\hfill
\begin{minipage}[t]{6cm}
\begin{center}
\includegraphics[width=6cm,height=6cm]{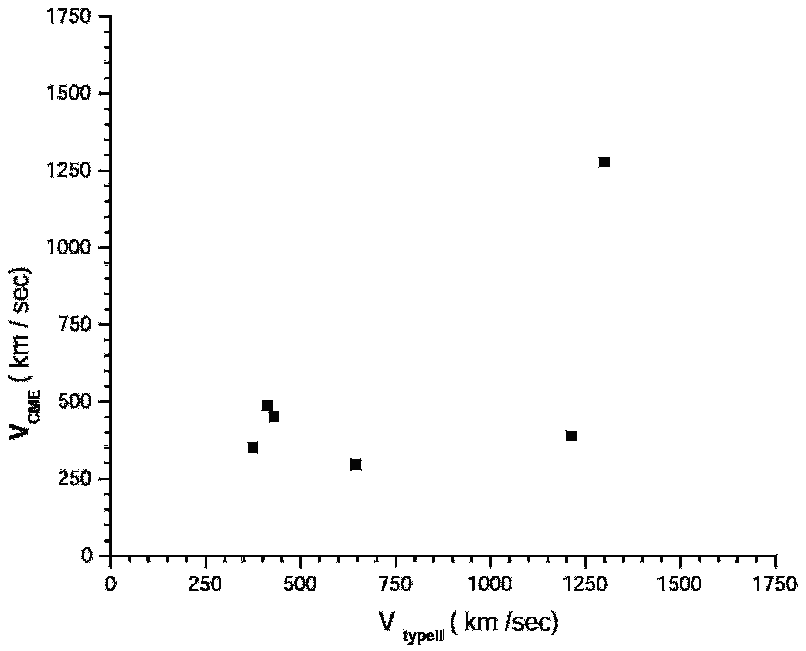}
\end{center}
\end{minipage}
 \caption{Left panel: Type II radial velocity versus SXR integrated Flux. The CME driven type II have
radial velocities approximately equal (within a factor of order 2 \cite{Classen}, cf. also next
panel) with the associated CME and appear to increase with SXR integrated Flux; the flare blast initiated type II
do not present any systematic dependence on the SXR flux. 
Right Panel: CME Velocity versus Type II Velocity; the two velocities are within a factor of
order 2 to each other.}
\label{Speeds}
\end{figure}

The relationship of the following flare--type II--CME parameters have been studied:

\begin{itemize}
\item{Flare average temperature: In \cite{Thomas} and \cite{Garcia} a method of 
calculation of the flare average temperature 
(or colour temperature) using the two channel fluxes from the GOES SXR detectors is demonstrated. 
We have obtained the flare temperature for each of the 11 SXR events of out data set following this method.
In figure \ref{FlareT} we plot the average temperature (Ta) as a function of SXR integrated flux for 
these events.}

\item{Rise, decay and Duration of SXR flares: These were obtained from the GOES profiles; in
figure \ref{FlareProperties}, left panel, we have plotted the flare duration versus the GOES SXR integrated flux,
in the right panel of the same figure we present a comparison of flare rise time versus decay time.}

\item{Velocities of type II events and CMEs: The type II radial velocities were obtained from 
\cite{Caroubalos04}, the CMEs linear speeds from the SOHO/LASCO catalogues on line. 
In figure \ref{Speeds} left panel we exhibit the type II radial velocity versus SXR integrated flux;
in the right panel the CME Velocity as a function of the type II Velocity. }

\end{itemize}

\section{Reasults, Discussion \& Conclusions}
We have analysed a relatively small group of complex events including type II shocks and SXR flares. 
A subgroup of these events was CME associated while the remaining were Non--CME events.
We have studied  the association of a number of characteristic parameters for each of the two subgroups, 
in an attempt to isolate features, or combinations thereoff, which might clearly distinguish them. The 
results are summarised as follows:
\begin{itemize}
\item{Flare average temperature: Our results in figure \ref{FlareT}
do not reveal a clear trend in the relationship of flare average temperature (Ta) versus SXR integrated flux;
in \cite{Kay} the CME--events have systematically lower tempreratures than Non--CME events
for the same peak intensity.}

\item{Rise, decay and Duration of SXR flares: With respect to flare duration (D) versus SXR integrated flux (F) 
the CME--events tend to have longer duration than Non--CME flares (cf. figure \ref{FlareProperties}) 
in accordance with previous results in \cite{Harrison} and \cite{Kay} although the latter derive these 
results comparing duration to peak intensity. As regards Flare rise time versus decay time 
no systematic difference appears for the two subgroups in accordance with similar results in \cite{Kay}.}

\item{Velocities of type II events and CMEs: As is demonstrated in figure \ref{Speeds} (right) the
type II and the CME velocities are within a factor of order 2 to each other, as expected from 
CME piston--driven shocks for the CME--associated subgroup (\cite{Classen}). This justifies the 
type II velocity relationship to the SXR integrated flux in figure \ref{Speeds} (left). 
In this panel the CME driven type II have
radial velocities approximately equal (within a factor of order 2 \cite{Classen}, 
with the associated CME and appear thus to increase with SXR integrated Flux as is expected for the 
CME speeds (\cite{Moon}, \cite{Caroubalos04}); 
the flare blast initiated type II do not present any systematic dependence 
on the SXR flux as they are expected to be affected, mostly, by the ambient corona conditions (\cite{Mann}).}

\end{itemize}

An original comparison of the CME--associted and the Non--CME subgroups does not demonstrate a 
substantial difference in the characteristics between them 
but for a tendency of the latter group to be shorter in duration than the former. 
Further work, with the use of a much larger sample is in progress.

\end{document}